\documentclass[twocolumn,secnumarabic,amssymb, nobibnotes, aps, prm]{revtex4-2}

\usepackage{graphicx}
\usepackage{xcolor}
\usepackage[english]{babel} 

\usepackage{natbib}
\setcitestyle{square, comma, sort }
\begin{document}

\title{Structural and Superconducting Properties of Ultrathin Ir Films on Nb(110)}%

\begin{abstract}
The ongoing quest for unambiguous signatures of topological superconductivity and Majorana modes in magnet-superconductor hybrid systems creates a high demand for suitable superconducting substrates. Materials that incorporate $s$-wave superconductivity with a wide energy gap, large spin-orbit coupling, and high surface quality, which enable the atom-by-atom construction of magnetic nanostructures using the tip of a scanning tunneling microscope, are particularly desired. Since single materials rarely fulfill all these requirements we propose and demonstrate the growth of thin films of a high-Z metal, Ir, on a surface of the elemental superconductor with the largest energy gap, Nb. We find a strained Ir(110)/Nb(110)-oriented superlattice for one to two atomic layer thin films, which transitions to a compressed Ir(111) surface for 10 atomic layer thick films. Using tunneling spectroscopy we observe proximity-induced superconductivity in the latter Ir(111) film with a hard gap $\Delta$ that is $85.3\%$ of that of bare Nb(110).
\end{abstract}

\author{Philip Beck}%
%\email{phbeck@physnet.uni-hamburg.de}
\affiliation{Department of Physics, University of Hamburg, Hamburg, Germany}
\author{Lucas Schneider}%
\affiliation{Department of Physics, University of Hamburg, Hamburg, Germany}
\author{Lydia Bachmann}%
\affiliation{Department of Physics, University of Hamburg, Hamburg, Germany}
\author{Jens Wiebe}
\email{jwiebe@physnet.uni-hamburg.de}
\affiliation{Department of Physics, University of Hamburg, Hamburg, Germany}
\author{Roland Wiesendanger}%
\affiliation{Department of Physics, University of Hamburg, Hamburg, Germany}
\date{November 2021}%
\maketitle
%\tableofcontents

\section{Introduction}
\label{sec:intro}
Magnet-superconductor hybrid (MSH) systems are at the heart of latest research efforts to realize topological superconductivity (TSC) and Majorana modes (MMs), which are candidates for topological qubits\cite{Nayak2008}. There are proposals to host TSC and MMs in multiple types of MSH systems, where the need of a static magnetic field to break time-reversal symmetry is lifted by magnetic components such as one-dimensional adatom chains forming a spin spiral\cite{NadjPerge2013, Kjaergaard2012}, ferromagnetic chains exposed to high spin-orbit coupling (SOC)\cite{Li2014}, two-dimensional ferromagnetic islands\cite{Roentynen2015, Li2016} and non-collinear magnetic films\cite{Nakosai2013, Chen2015}, which are proximity-coupled to $s$-wave superconductors. 
\newline
Indeed, first experiments using a scanning tunneling microscope (STM) showed that self-assembled ferromagnetic Fe chains on Pb(110)\cite{NadjPerge2014, Ruby2015, Jeon2017} and chains of Fe on Re(0001) fabricated by tip-induced atom manipulation in a spin helix ground state\cite{Kim2018} display spectroscopic characteristics compatible with MMs. However, for those systems, the observation of MMs in a hard topological gap, where the density of states approaches zero, is still to be demonstrated. Challenges are a complex adsorption geometry and intermixing effects in the Fe/Pb(110) system and the resolveability of in-gap features in scanning tunneling spectroscopy (STS) of the Fe/Re(0001) system which suffers from the small superconducting gap of Re\cite{Kim2018, Schneider2020}. Therefore, more recent work has focused on employing Nb(110) as a substrate for Fe\cite{Friedrich2021, Crawford2021} and Mn\cite{Schneider2021, Schneider2021a} adatoms where in-gap features become clearly observable and distinguishable due to the larger gap and higher critical temperature $T_\mathrm{C}=9.2~\mathrm{K}$ compared to Re. While it was found that considerable SOC is present in Nb based MSH systems\cite{Beck2021}, it turns out to be too low to gap out all Shiba bands beyond the experimental resolution. Further, only collinear magnetic ground states were calculated\cite{Laszloffy2021} and experimentally observed\cite{Schneider2021b} for most Mn chains on Nb(110), resulting in non-ideal circumstances to realize MMs protected by a hard topological gap.
\newline
The goal of this study is to fabricate a superconducting heterostructure consisting of a thin film of a high-Z metal grown on the (110) surface of a Nb single crystal. The latter is intended to provide the excellent superconducting properties, while the former serves to increase SOC and the Dzyaloshinsky-Moriya interaction\cite{Dzyaloshinsky1958, Moriya1960} within the nanostructure assembled on top, which favors non-collinear magnetic ground states. Superconductivity will be proximity-induced into the high-Z metal overlayer, if the film's thickness is kept low enough\cite{Belzig1996, Reeg2016,Tomanic_2016}. Furthermore, the spin-carrying states of the magnetic nanostructure will hybridize with the electronic states of the high-Z metal, which should induce strong SOC in the nanostructure, as SOC roughly scales with the atomic number Z of the species. Therefore, such systems should be ideal to realize TSC in one- and two-dimensional magnetic nanostructures. Ir is a particularly well-suited candidate for the high-Z metal overlayer material, since Ir(111) surfaces have already been shown to serve as templates for the realization of spin-spiral and skyrmion phases\cite{Bergmann2007, Heinze2011, Romming2015}.   
\newline
Here we present our growth study of Ir on Nb(110) using STM and low-energy electron diffraction (LEED) to determine the crystal structure of Ir thin films. We reveal a structural transition with increasing layer thickness from an Ir(110)-like rectangular unit cell for few-layer samples to the desired Ir(111) surface structure for thicker Ir films. After obtaining the desired surface structure of the Ir film, we probe the proximity-induced superconductivity using STS and find a fully pronounced superconducting gap, which still is $85.3\%$ in size of that of bare Nb(110). Therefore, our results pave the way for studying one- and two-dimensional magnetic nanostructures such as spin spirals and skyrmions with proximity-induced superconductivity, where the superconducting properties are close to that of bare Nb(110) enabling a sufficient energy resolution of tunneling spectroscopy in order to disentangle trivial and topological states.
\begin{figure*}
	
	\includegraphics[scale=1]{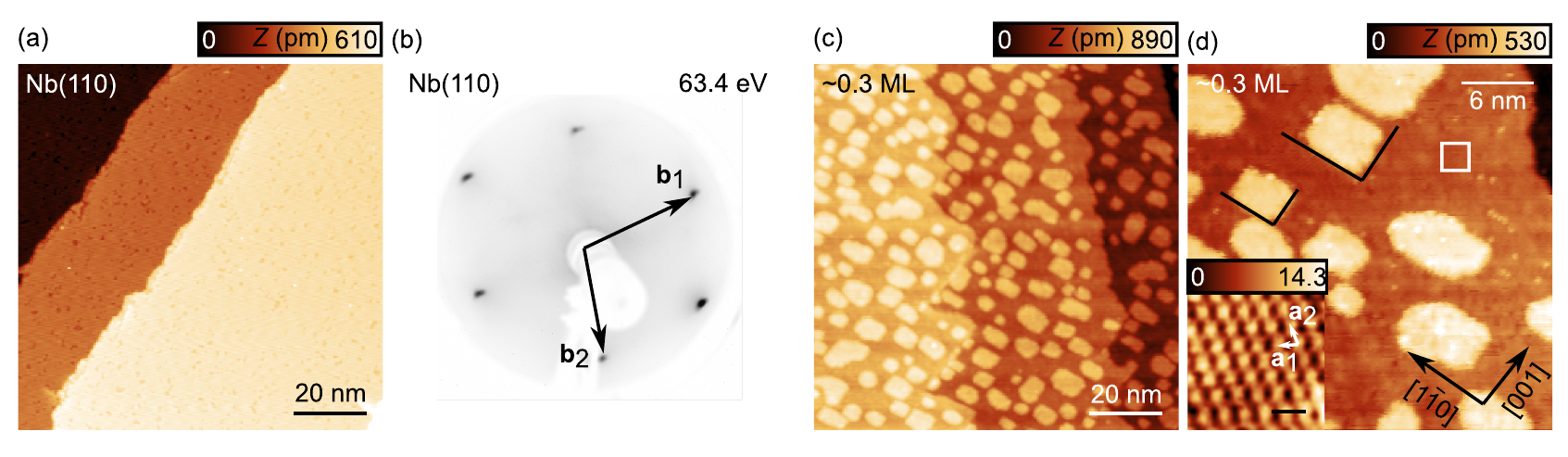}
	\caption{\label{fig:SubML}(a) STM image  and (b) LEED pattern showing the surface quality of clean Nb(110) after the flashing process. The LEED pattern was obtained with a beam energy of $63.4~\mathrm{eV}$. Black arrows and labels denote reciprocal lattice vectors. (c)-(d) Overview STM images of an Ir/Nb(110) sample with a coverage of $\sim 0.3 ~\mathrm{ML}$. Black arrows and labels denote crystallographic directions valid for panels (c) and (d). An atomic resolution STM image taken at the location of the white box is shown as an inset in panel (d). The black scale bar corresponds to a length of $500~\mathrm{pm}$. A Gaussian filter was applied in order to highlight the atomic resolution. White arrows and labels mark lattice vectors ($\mathbf{a}_1$ and $\mathbf{a}_2$) of the clean Nb(110) substrate. Measurement parameters: (a) $V_{\mathrm{bias}}= -1~\mathrm{V}, I= 1~\mathrm{nA}$, (c),(d) $V_{\mathrm{bias}}= -1~\mathrm{V}, I= 1.15~\mathrm{nA}$, inset of (d) $V_{\mathrm{bias}}= -10~\mathrm{mV}, I= 5~\mathrm{nA}$.}
\end{figure*}

\section{Methods}
\label{sec:methods}
The study was performed in two different ultra-high vacuum (UHV) systems. The samples with Ir coverages below 1 monolayer (ML), as well as at 1 ML and at 2 MLs coverages (Figures~\ref{fig:SubML} and \ref{fig:MLGrowth}) were prepared and investigated in a commercial Omicron UHV chamber with a base pressure of $p \approx 1 \times 10^{-11}~\mathrm{mbar}$ equipped with a home-built variable-temperature (VT)-STM similar to the one described in \cite{Kuck2008} and a system for LEED measurements. The STM was cooled to 30 K using a flow-cryostat operated with liquid helium. An electrochemically sharpened W tip was used in this set-up.
\newline
The sample with an Ir coverage of $\sim 10$ ML (Figure~\ref{fig:TenML}) was investigated in a home-built UHV system equipped with a STM operated at a temperature of $320~\mathrm{mK}$ \cite{Wiebe2004}. A mechanically sharpened Nb tip was used for the measurements in this system.
\newline
Clean Nb(110) was prepared by high temperature flashes\cite{Odobesko2019} using home-built e-beam stages, reaching temperatures above $2400^{\circ}$C. Ir was deposited on freshly flashed samples using commercial e-beam evaporators (EFM3 by Focus) equipped with an Ir rod ($99.9 \% $ purity) as evaporant. After Ir deposition the samples were post-annealed on the e-beam stage for approximately two minutes at a power of $2~\mathrm{W}$.
\newline
The coverage for films thicker than 1 ML was calculated by calibration measurements: We measured the evaporator flux and the evaporation time for a calibration sample and subsequently analyzed the Ir coverage of the Nb(110) sample ($<1$ ML) using STM. The determined evaporation rate was then used to calculate an estimated coverage of later sample preparations. Note that this method assumes a pseudomorphic ML and doesn't account for structural changes in the Ir film. Thus, it should be seen as an estimate for coverages larger than 2 MLs.    

\section{Results}
An STM image of the typical surface quality of the Nb(110) single crystal obtained after high temperature flashes, which are required to deplete the crystal of oxygen, is shown in Figure~\ref{fig:SubML}(a). As recently reported and visible in the STM image, this preparation results in largely clean Nb(110) terraces\cite{Odobesko2019}. Only minor oxygen impurities remain and are visible as apparent depressions. A LEED pattern measured on this sample is shown in Figure~\ref{fig:SubML}(b), where one can see sharp diffraction spots, whose symmetry matches that of a body-centered cubic, bcc, (110) surface. This highlights the high quality of the as-prepared sample. Reciprocal lattice vectors are marked by black arrows and are labeled $\mathbf{b}_1$ and $\mathbf{b}_2$. As apparent from the large clean regions of Nb(110) and the low concentration of oxygen defects, the sample quality is sufficient to study the growth of Ir in the clean limit.

\subsection{Sub-ML coverage}
A sample with an iridium coverage of $\sim0.3~ \mathrm{ML}$ is shown in Figures~\ref{fig:SubML}(c) and (d), which are large-scale overview STM images highlighting the quality and homogeneity of the Ir islands. As apparent from the irregularly shaped step edges in Figure~\ref{fig:SubML}(c), in comparison to those of bare Nb(110) in Figure~\ref{fig:SubML}(a), we conclude that iridium grows via the step-edge decoration and step-flow, as well as in a  free-standing island mode. Both, the islands and the substrate, appear to be atomically flat, with the exception of few randomly distributed point-like defects. Furthermore, the Ir islands mostly have approximately rectangular shapes, whose edges are preferably oriented along the directions indicated by black lines in Figure~\ref{fig:SubML}(d). The atomic resolution image taken on the substrate (inset in the bottom left corner of \mbox{Figure}~\ref{fig:SubML}(d)) shows the atomic structure of bare Nb(110), i.e. a centered rectangular unit cell with edges along the $[1\overline{1}0]$- and the $[001]$-direction as shown in Figure~\ref{fig:SubML}(d). The fact that large areas of clean Nb(110) are still found, even after Ir evaporation and post-annealing processes, is crucial to highlight the Ir/Nb(110) interface quality. Obviously, oxygen diffusion from the bulk crystal to the surface or impurities from the evaporator beam can be largely excluded, also for the further preparations as long as the post-annealing protocol is not altered. Further, a comparison of the crystallographic directions marked in Figure~\ref{fig:SubML}(d) and the preferred edge orientation of the iridium islands highlighted in \mbox{Figure}~\ref{fig:SubML}(d) hints towards  a rectangular crystal structure of the Ir islands. Obtaining clear atomic resolution images of the Ir islands was not possible in this experiment.
\begin{figure*}
	
	\includegraphics[scale=1]{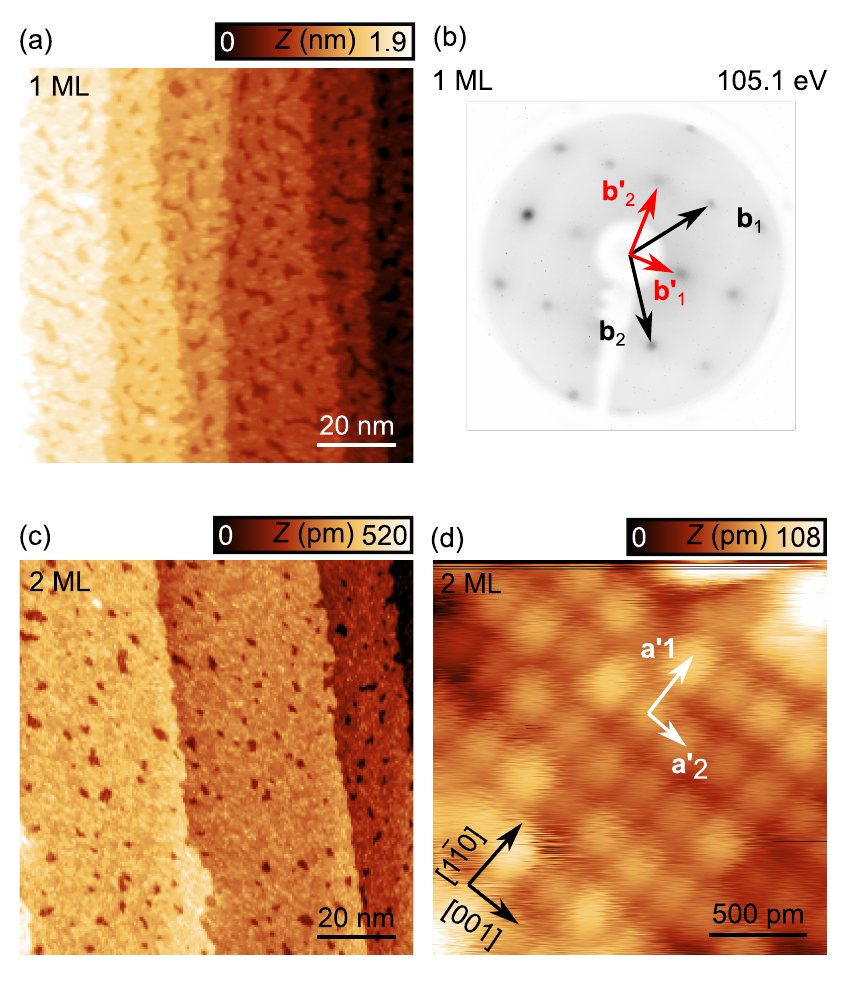}
	\caption{\label{fig:MLGrowth}(a) STM image ($V_{\mathrm{bias}}= -1~\mathrm{V}, I= 20~\mathrm{pA}$) and (b) LEED pattern characterizing the growth of 1 ML Ir on clean Nb(110). The LEED pattern was obtained with a beam energy of $105.1~\mathrm{eV}$. Black arrows and labels in (b) denote reciprocal lattice vectors, which were observed for clean Nb(110). Red ones mark LEED spots observed only after the growth of 1 ML Ir. (c)-(d) STM images of an Ir/Nb(110) sample with a coverage of $\sim 2 ~\mathrm{MLs}$. (d) Black arrows denote crystallographic directions as obtained from Figure~\ref{fig:SubML}(d) and white arrows mark the lattice vectors ($\mathbf{a}^{'}_1$ and $\mathbf{a}^{'}_2$) determined from the visible superstructure. (Measurement paramters: (c) $V_{\mathrm{bias}}= -1~\mathrm{V}, I= 20~\mathrm{pA}$ and (d) $V_{\mathrm{bias}}= -20~\mathrm{mV}, I= 8~\mathrm{nA}$).}
\end{figure*}

\subsection{1 \& 2 MLs - Ultrathin Film Limit }
As we have established that Ir grows in an ordered and flat fashion, and that the Ir/Nb(110) interface remains clean, even after applying post-annealing, we now proceed to study samples with higher coverage of Ir, which are needed for the purpose of single atom experiments or for the growth of metallic transition metal thin films onto Ir thin films. Figure~\ref{fig:MLGrowth}(a) shows a large-scale STM image obtained for a sample with a coverage of 1 ML Ir. From Figure~\ref{fig:MLGrowth}(a) it becomes apparent, that the first ML isn't fully closed, as there are still holes in the film. Further, irregularly shaped step edges indicate that the formation of a second ML might have begun. Flat connected areas are visible between the holes of the first ML, which are only disrupted by point-like defects that have been observed for samples with sub-ML coverage (see Figure~\ref{fig:SubML}(d)) as well. A LEED pattern measured on this sample is shown in Figure~\ref{fig:MLGrowth}(b). Using this data, a model of the crystal structure can be derived. To start with, the LEED spots of bare Nb(110) (c.f. Fig.~\ref{fig:SubML}(b)) are still found in this measurement and are marked by black arrows which are labeled by $\mathbf{b}_1$ and $\mathbf{b}_2$. Additional sharp and bright spots are observable at locations marked by red arrows, which are labeled $\mathbf{b}^{'}_1$ and $\mathbf{b}^{'}_2$. Since the emergence of these new LEED spots is linked to a coverage of the Nb(110) substrate with an ultrathin Ir film, we can derive its crystal structure from them. The newly observed LEED spots, marked by $\mathbf{b}^{'}_1$ and $\mathbf{b}^{'}_2$, lie directly in between $\mathbf{b}_1$ and $\mathbf{b}_2$ or $\mathbf{b}_1$ and $- \mathbf{b}_2$, respectively. Therefore, they can be described as linear combinations of the reciprocal lattice vectors of Nb(110)
\newline
\begin{equation}
	\label{eq:b_1}
	\mathbf{b}^{'}_1 =\frac{\mathbf{b}_1}{2}+ \frac{\mathbf{b}_2}{2} 
\end{equation}
\newline
and
\begin{equation}
	\label{eq:b_2}
	\mathbf{b}^{'}_2 =\frac{\mathbf{b}_1}{2}- \frac{\mathbf{b}_2}{2}
\end{equation}
which enables a calculation of the real-space lattice vectors of this structure. Using reciprocal lattice vectors of the bcc(110) Nb surface for $\mathbf{b}_1$ and $\mathbf{b}_2$ and inserting them into Equations~(\ref{eq:b_1}) and (\ref{eq:b_2}), we calculate the real-space lattice vectors of the Ir film to be
\newline
\begin{equation}
	\label{eq:a_1}
	\mathbf{a}^{'}_1 =\left(
	\begin{array}{c}
		0\\
		\sqrt{2}\\
	\end{array}
	\right) \cdot a  
\end{equation}
\newline
and
\begin{equation}
	\label{eq:a_2}
	\mathbf{a}^{'}_2 = \left(
	\begin{array}{c}
		1\\
		0\\
	\end{array}
	\right) \cdot a . 
\end{equation}
Here, $x$ and $y$ correspond to the $[001]$ and $[1\overline{1}0]$-direction, respectively, and $a$ is the lattice constant of Nb(110) ($\sim 330~\mathrm{pm}$) as determined from the atomic resolution image in the inset of Figure~\ref{fig:SubML}(d). Therefore, we find a rectangular unit cell with lattice constants of $	|\mathbf{a}^{'}_1|=467~\mathrm{pm}$ and $	|\mathbf{a}^{'}_2|=330~\mathrm{pm}$ for the first ML of Ir, on the centered rectangular unit cell of bare Nb(110). We speculate that the crystal structure observed for the first ML of Ir on Nb(110) was also present in the islands of Figure~\ref{fig:SubML} due to the predominant rectangular shape of the islands.
\newline
\begin{figure*}
	
	\includegraphics[scale=1]{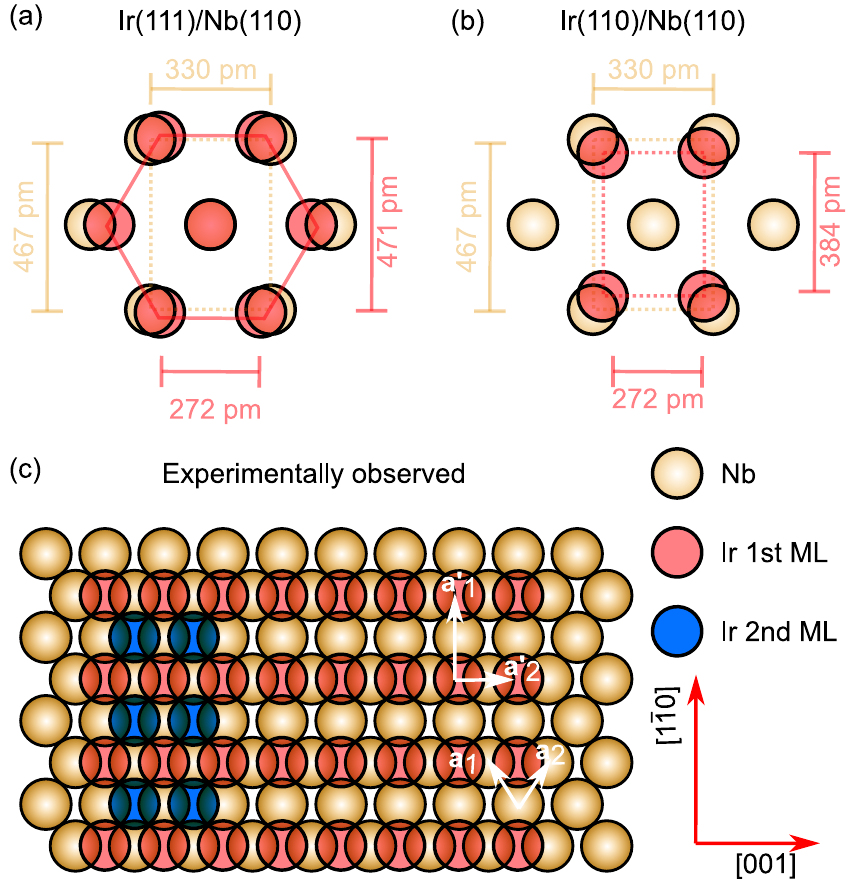}
	\caption{\label{fig:MLModel}(a)-(b) Ball models of two different possible crystallographic orientations for the growth of Ir on Nb(110). Yellow and pink spheres represent niobium and iridium atoms, respectively. (a) shows the unrelaxed hexagonal unit cell of Ir(111), and (b) the unrelaxed rectangular one of Ir(110) on top of the unrelaxed pseudo-hexagonal unit cell of Nb(110). (c) Structural model for 1 and 2 MLs of Ir on Nb(110) suggested by the experimentally observed STM images and LEED patterns in Fig.~\ref{fig:MLGrowth}. Blue spheres correspond to Ir atoms forming the 2nd ML. White arrows indicate the lattice vectors of the Nb(110) substrate ($\mathbf{a}_1$ and $\mathbf{a}_2$) and of the 1st ML Ir ($\mathbf{a}^{'}_1$ and $\mathbf{a}^{'}_2$). Red arrows in the bottom right corner mark crystallographic directions, valid for all three panels.}
\end{figure*}
Subsequently, we prepared samples with an Ir coverage of 2 MLs, to determine whether they grow in a similar crystal structure. From the large scale images in Figure~\ref{fig:MLGrowth}(c), we find that the second ML is not fully closed either, as there are holes with atomic step height. However, in contrast to the sample with 1 ML coverage (Figure~\ref{fig:MLGrowth}(a)), we find that the step edges are straight which might indicate the absence of step-flow growth of a third ML.
\newline
The atomic resolution image shown in Figure~\ref{fig:MLGrowth}(d), taken on a clean part of the terrace, clearly shows a simple rectangular lattice, whose lattice vectors are indicated by white arrows and labels. Black arrows highlight the crystallographic directions of the underlying Nb(110) lattice as obtained from Figure~\ref{fig:SubML}(d). Analyzing the 2D-FFT of this atomic resolution image (not shown) confirms the existence of a simple rectangular lattice and yields absolute values for $|\mathbf{a}^{'}_1|=480~\mathrm{pm}$ and $|\mathbf{a}^{'}_2|=340~\mathrm{pm}$ which are in very good agreement with the values calculated from the LEED pattern of the Ir ML in Figure~\ref{fig:MLGrowth}(b). We conclude that the 2nd ML growth occurs in the same crystal structure as the 1st ML of Ir. 
\newline 
As discussed in Section~\ref{sec:intro}, it would be desirable to obtain an Ir film with a crystal structure similar to that of an Ir(111) surface. The mismatches of the Ir(111) and Ir(110) surfaces as grown on Nb(110) are illustrated in Figures~\ref{fig:MLModel}(a) and (b), respectively. The literature values of the lattice constants for bulk bcc Nb ($330~\mathrm{pm}$) and face-centered cubic (fcc) Ir ($384~\mathrm{pm}$)\cite{Ashcroft2012} were used to calculate the nearest neighbor distances in $[001]_{\mathrm{Nb}}$- and $[1\overline{1}0]_{\mathrm{Nb}}$-directions.
\newline
A comparison of the unrelaxed crystal structures in the orientation Ir(111)/Nb(110), as illustrated in Figure~\ref{fig:MLModel}(a), shows that there is a very small mismatch $(a_{\mathrm{Ir}}-a_{\mathrm{Nb}} ) / a_{\mathrm{Ir}}$ in $[1\overline{1}0]$-direction of only $0.8\%$. However, the mismatch in $[001]$-direction, $-21.3\%$, is comparably large. While the mismatch in the $[001]$-direction remains the same for the growth orientation Ir(110)/Nb(110) shown in Figure~\ref{fig:MLModel}(b), it is increased to the similar value, $-21.6\%$, in $[1\overline{1}0]$-direction.  
\newline
As suggested by the experimentally measured crystal structures and lattice parameters of the ultrathin Ir films, sub-ML-, one ML- and two MLs of Ir most probably grow in a  Ir(110)/Nb(110)-oriented superlattice. If we assume, that the Ir atoms prefer a four-fold coordinated hollow adsorption site on the Nb(110) surface, we end up with the structural model illustrated in Figure~\ref{fig:MLModel}(c). The resulting tensile strains $\epsilon=(a_{\mathrm{exp.}}-a_{\mathrm{Ir(110)}} ) / a_{\mathrm{Ir(110)}}$ of the Ir layer along the $[001]$- and $[1\overline{1}0]$-directions are $21.3\%$ and $21.6\%$, compared to the unrelaxed Ir(110) surface.
\newline
\begin{figure*}
	
	\includegraphics[scale=1]{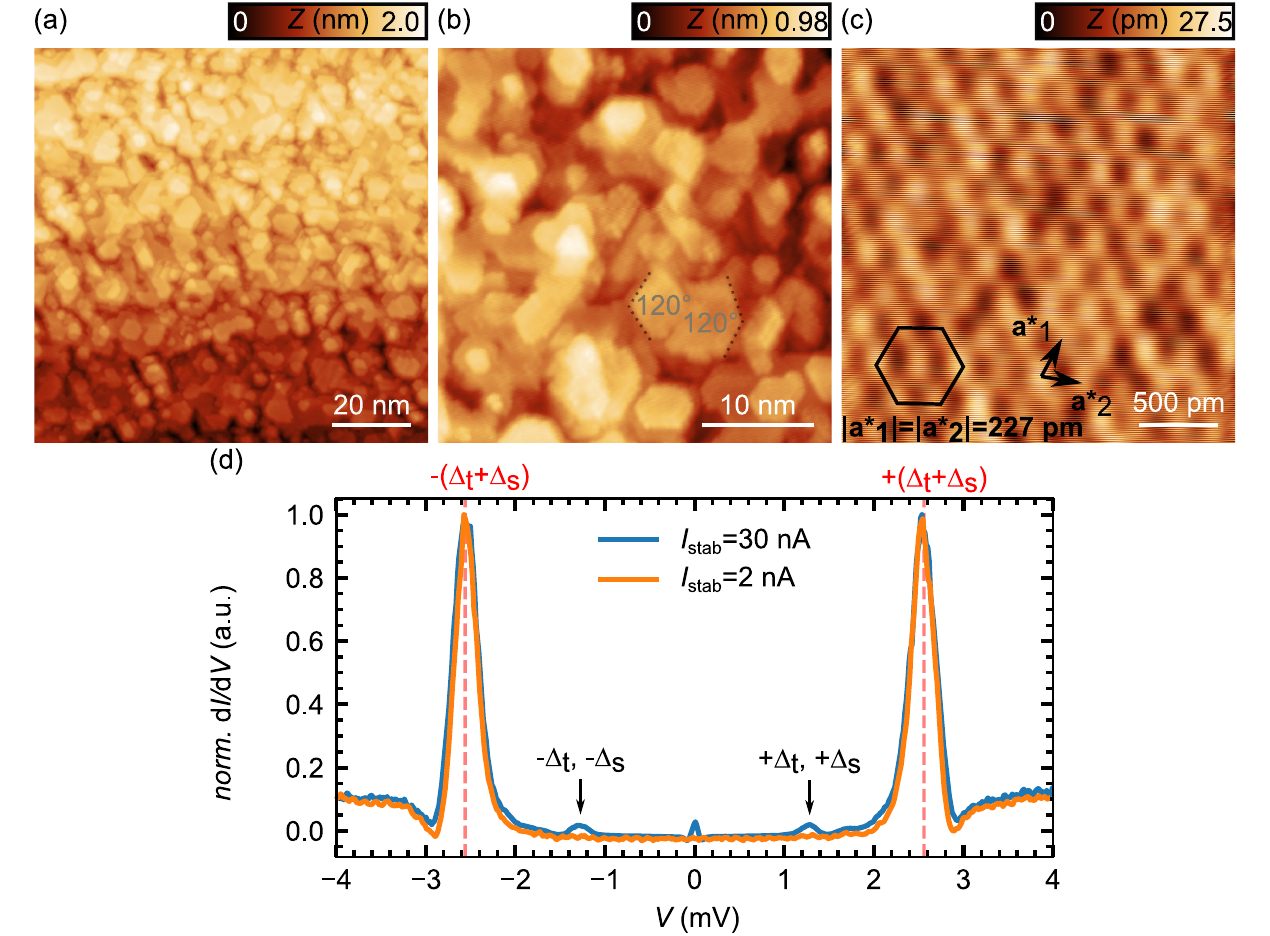}
	\caption{\label{fig:TenML} (a)-(b) Overview STM images of a sample with $\sim 10$ MLs of Ir coverage. Dashed lines mark common angles enclosed by the edges of Ir islands. (c) Atomic resolution STM image displaying a hexagonal lattice as indicated by the black hexagon marking a conventional unit cell of the lattice. Black arrows and labels denote the lattice vectors $\mathbf{a}^{*}_1$ and $\mathbf{a}^{*}_2$. (d) $\mathrm{d}I/\mathrm{d}V$ spectrum taken on an Ir island, using a bulk Nb tip at two different stabilization currents. The spectra are normalized by their respective maximum value. Black and red labels mark the superconducting coherence peaks at $\pm(\Delta_\mathrm{t} + \Delta_\mathrm{s})$ and the merged MAR's at $\pm \Delta_\mathrm{t}$ and $\pm \Delta_\mathrm{s}$. Measurement parameters: (a) $V_{\mathrm{bias}}= -1~\mathrm{V}, I= 200~\mathrm{pA}$, (b) $V_{\mathrm{bias}}= -10~\mathrm{mV}, I= 2~\mathrm{nA}$, (c) $V_{\mathrm{bias}}= -10~\mathrm{mV}, I= 1~\mathrm{nA}$, (d)  $V_{\mathrm{stab}}= -4~\mathrm{mV},  I_{\mathrm{stab}}= 2~\mathrm{nA}$ for the orange curve and $I_{\mathrm{stab}}= 30~\mathrm{nA}$ for the blue curve.}
\end{figure*}
\subsection{Ten MLs - Thin Film Limit}
\label{sec:10ML}
To eventually achieve a hexagonal surface symmetry with lattice vectors similar to that of Ir(111), we evaporated Ir with a higher coverage. Since Ir(111) is a common stable surface of an Ir single crystal, we would expect that the crystal structure of films on Nb(110) will eventually relax into this hexagonal crystal surface with increasing film thickness, once the strain energy accumulated in the first Ir layers exceeds that of the Ir/Nb interface energy. We prepared a sample with a coverage of 10 MLs to investigate if there is a structural transition for higher coverages.
\newline 
The resulting sample is shown in Figures~\ref{fig:TenML}(a)-(c), where (a) and (b) are large-scale STM images and (c) is an atomic resolution image. As for the thin films investigated before, the Ir does not grow perfectly layer by layer for the used post annealing power and time. One rather finds islands, which are largely hexagonal in shape, have sharp edges, mostly show $120^\circ$ corners (marked in Figure~\ref{fig:TenML}(b)) and are atomically flat and clean. As seen from the atomic resolution image in Figure~\ref{fig:TenML}(c), these islands now indeed have a hexagonal surface structure. A conventional unit cell as well as the lattice vectors $\mathbf{a}^{*}_1$ and $\mathbf{a}^{*}_2$ of this hexagonal structure are marked by black arrows and labels. The absolute length of the direct lattice vectors may be calculated to be $|\mathbf{a}^{*}_1| =|\mathbf{a}^{*}_2|=227~\mathrm{pm}$. The ideal value for a perfect Ir(111) surface would be $\sim 272~\mathrm{pm}$ (see Figure~\ref{fig:MLModel}(a)), which results in a compressive strain of $-16.5\%$. 
\newline
Apart from the crystal structure, another crucial physical property of the Ir films, which is required for the experiments proposed in Section~\ref{sec:intro}, is superconductivity. Two important criteria to judge the proximity-induced superconductivity in the Ir film are the size of the superconducting gap $\Delta_\mathrm{s}$ and whether it stays a fully pronounced "hard" gap, i.e. $\mathrm{d}I/\mathrm{d}V \sim \mathrm{DoS}$ approaches zero inside the superconductor's gap.
\newline
To evaluate these criteria, two tunneling spectra obtained on the Ir film are shown in Figure~\ref{fig:TenML}(d). For one, a spectrum with a small tip-sample distance is shown in the blue curve. Apart from a typical superconducting energy gap with coherence peaks at $\pm |\Delta_\mathrm{s}+\Delta_\mathrm{t}|=\pm 2.56~\mathrm{mV}$, where $\Delta_\mathrm{t}$ is the superconducting gap of the tip, a Josephson peak becomes apparent at zero bias. Further in-gap states are observed at $\pm 1.28~\mathrm{mV}$, which we determine to be multiple Andreev reflections (MARs)\cite{Ternes2006}. MARs are expected to occur at $\pm |\Delta_\mathrm{s}|$ and $\pm |\Delta_\mathrm{t}|$. Since only one pair of MARs is observed, we conclude that both $\pm |\Delta_\mathrm{t}|$ and $\pm |\Delta_\mathrm{s}|$ are $\pm 1.28~\mathrm{meV}$ large, which matches precisely with the observation of coherence peaks at $\pm 2.56~\mathrm{mV}$. Compared to typical values of $\pm |\Delta_\mathrm{s}|$ for bare Nb(110) at $300~\mathrm{mK}$ ($1.50~\mathrm{meV}$\cite{Beck2021}), the proximity-induced superconducting energy gap is reduced by only $14.7\%$.
\newline
Additionally, a tunneling spectrum with common stabilization parameters of $V_{\mathrm{stab}}= -4~\mathrm{mV},  I_{\mathrm{stab}}= 2~\mathrm{nA}$ is shown in orange. A clear superconducting energy gap with coherence peaks at $\pm |\Delta_\mathrm{s}+\Delta_\mathrm{t}|=\pm 2.56~\mathrm{mV}$ is observed. The $\mathrm{d}I/\mathrm{d}V$ signal drops to zero between the coherence peaks and $\pm |\Delta_\mathrm{t}|=\pm 1.28~\mathrm{mV}$ indicating a hard superconducting gap.
\section{Discussion}
Thin films of Ir grown on clean Nb(110) undergo a crystal structure transition from an Ir(110)-oriented superlattice with a rectangular unit cell stretched by $\sim 21.5\%$ for a low coverage of one to two MLs, to an Ir(111)-oriented hexagonal crystal structure contracted by $-16.5\%$ for higher coverages of approximately ten MLs. We speculate, that the compression of the hexagonal lattice compared to bulk Ir(111) is a result of the structural transition from the Ir(110)-oriented superlattice to the Ir(111)-oriented structure: The lattice constants of the Ir(110)-oriented superlattice will approach the unrelaxed values indicated by red lines in Figure~\ref{fig:MLModel}(b) with increasing film thickness. The Ir(110) to Ir(111) transition could then occur by the occupation of the center of the rectangular unit cell by an additional Ir atom, resulting in a nearest-neighbour distance of $235~\mathrm{pm}$. A comparison to the experimentally extracted value of $227~\mathrm{pm}$ yields good agreement. The observed structural behavior differs from other known growth modes of fcc transition metals such as Ag\cite{Ruckman1988, Tomanic_2016}, Au\cite{Ruckman1988}, Pt\cite{Pan1987} and Pd\cite{Strongin1980} on Nb(110), which tend to form a pseudomorphic layer for one ML coverage, which transitions to the fcc(111) structure for a higher coverage. A recent growth study of Bi on clean and on reconstructed Nb(110) has found a deviant growth mode as well\cite{Boshuis2021}.
\newline
The observation of a strained Ir(110)-oriented superstructure with a rectangular unit cell for one and two MLs alone is surprising: the preparation of a clean and unreconstructed Ir(110) surface seems to be unachievable starting from an Ir single crystal, since the surface stabilizes by forming $(311)$ facets\cite{Koch1991, Kuntze1998}. Therefore, the growth of ultrathin films of Ir on Nb(110) enables the study of the Ir(110) surface, and could be of interest for further experiments. 
\newline
Regarding the application of Ir/Nb(110) in MSH systems, however, the sample with a coverage of 10 MLs displaying a hexagonal lattice is more interesting. The limited island size observed in Figures~\ref{fig:TenML}(a) and (b) might be explained by the strain of the hexagonal lattice, compared to bulk Ir(111), which we expect to further approach the ideal value with increasing layer thickness. We, thus, speculate that one could increase the island's size and eventually achieve continuous layers by growing thicker Ir films. However, it should be noted that the proximity-induced superconductivity is already affected by the thickness of the Ir layers studied here, as the gap is reduced by $14.7\%$ compared to bare Nb(110). Therefore, it would be useful to further study the influence of the Ir layer thickness on the proximity-induced superconductivity, to allow for the preparation of a sample in the sweet spot of reaching the optimal lattice constant of Ir(111) with large terraces while maintaining the superconducting properties of Nb(110) as much as possible. 
\section{Conclusion}
In conclusion, we grew thin films of Ir on Nb(110) at various coverages and determined the structural properties layer-dependently. While we find a strained Ir(110)-oriented superstructure for ultrathin films (1-2 MLs), we obtain the highly desired Ir(111) surface structure at a higher coverage (10 MLs). We determined the proximity-induced superconductivity in this sample and observed a fully pronounced, hard gap, which only differs from that of Nb(110) by its slightly reduced size. Therefore, we pave the way to deposit multiple highly interesting magnetic transition metals in the form of single atoms or thin films on this surface and thereby study the interplay of non-collinear magnetism or magnetic materials subject to high SOC with superconductivity.  
\section*{Acknowledgments}
P.B., R.W., and J.W. gratefully acknowledge funding by the Deutsche Forschungsgemeinschaft (DFG, German Research Foundation) – SFB-925 – project 170620586. L.S., R.W., and J.W. gratefully acknowledge funding by the Cluster of Excellence 'Advanced Imaging of Matter' (EXC 2056 - project ID 390715994) of the DFG. R.W. gratefully acknowledges funding of the European Union via the ERC Advanced Grant ADMIRE (grant No. 786020). We acknowledge fruitful discussions with W. Li.  
%\bibliography{20210511_v2}


\end{document}